\documentclass[11pt,twoside]{article}
 
\usepackage{asp2006} 
\usepackage{epsf} 
\usepackage{lscape} 
\usepackage{graphicx} 
 
\begin{document} 

\title{Dust from AGB stars}   %%% Fill in title 
\author{Anja C.\,Andersen}   %%% Fill in author names 
\affil{Dark Cosmology Centre, Niels Bohr Institute, University of Copenhagen, Juliane Maries Vej 30, DK-2100 Copenhagen, Denmark. Email: anja@dark-cosmology.dk}    
 
\begin{abstract} %%% Abstract to run on from here. 
Dust is formed in the expanding atmosphere during late stages of 
stellar evolution.  
Dust influences the dynamics and 
thermodynamics of the stellar atmosphere by its opacity. 
The dust opacity depends both on the optical properties of 
the grain material as well as on the amount of dust present. 
A rich source of information on some mineral phases of dust
in AGB stars comes from the study of presolar grains from
meteorites. This paper presents a short overview of presolar
grains studies and describes how the optical properties of 
dust grains are obtained in the laboratory.  
\end{abstract}

%%%%%%%%%%%%%%%%%%%%%%%%%%%% 
\section{Dust and AGB stars} 
%%%%%%%%%%%%%%%%%%%%%%%%%%%% 
 
The formation of molecules and dust grains is strongly dependent on the time-dependent dynamical processes that affect the atmospheres of AGB stars, i.e.\ pulsations and propagating shock waves. 

Two basic condition must be fulfilled in the AGB star atmosphere for dust formation to succeed, relatively low temperature ($<$ 2000~K) combined with high density. The sufficiently low temperature is present above the photosphere. As a consequence of the pulsation, the outer atmosphere is levitated, resulting in a temporary reservoir of relatively dense gas at a certain distance from the photosphere, and thus increasing the efficiency of dust formation. 
 
Observations indicate that dust is a significant ingredient of the outflows from AGB stars. It has therefore long been assumed that the mass loss from AGB stars is driven by radiation pressure on dust grains \citep[e.g.][]{HerasHony05}, however, recent frequency-dependent numerical models of M-type stars have not yet been able to support this (see H{\"o}fner and Woitke in these proceedings).  
 
In this paper the focus is on what we know about the dust properties from studies of presolar AGB dust grains from meteorites, and from laboratory studies of dust analogues.  
 
%%%%%%%%%%%%%%%%% 

\section{Dust from AGB stars - Presolar grains extracted from meteorites} 
%%%%%%%%%%%%%%%% 
 
While most of the material that went into the making of the Solar System 
was throughly processed and mixed, thus losing isotopic heterogeneity 
and all memory of its origin, small quantities of refractory dust 
grains survived the events that led to the formation of the Solar System 
and such grains have been found in primitive meteorites  
\citep[Zinner 1998; Hoppe \& Zinner 2000; Nittler et al.\ 2003; Zinner 2004;][]{HoppeOtt07} \nocite{HoppeZinner00} \nocite{Nittler03} \nocite{Zinner04} \nocite{Zinner98}. 
 
When the matrix material from these primitive meteorites was heated in the laboratory, it was 
realized already in the early 1960'ies that at certain temperatures the matrix released noble gases with 
an isotopic composition markedly different from everything else in the Solar System.  
Although the noble gasses 
are far less abundant than condensable elements, they are present in 
measurable quantities in virtually all meteorites.   
Isotopic variations in noble gases are quite often orders of 
magnitudes larger than the isotopic variations in the rock-forming elements, 
because even the most primitive bulk meteorites contain only a small fraction 
($\approx 10^{-4}$ for Xe and $\approx 10^{-9}$ for He) of their Solar 
abundances.  This means that the noble gases are less contaminated by  
``normal'' matter, than other more abundant elements.   
It was quickly realized that in view of the volatility of these gases, 
they must be trapped in solid grains. 
The carrier should be a refractory mineral which would be 
chemically resistant under the 
conditions prevailing in the interstellar medium, during formation of the 
Solar System and evolution of the meteorite parent body.  Based on 
these requirements, two forms of matter can be expected to have 
survived and to be identifiable as presolar: refractory oxides and 
refractory carbon compounds, which are the chemically stable forms for 
different carbon to oxygen (C/O) ratios.   
For the Solar System ratio 
of C/O = 0.42  \citep{Anders_etal89}, the oxides, not the carbon 
compounds, are the stable phase. 
 
After years of trials with different chemical purification of the matrix 
material, and subsequent stepwise heating and isotopic noble gas measurements, 
the first presolar grains were finally isolated by \citet{Lewis_etal87}, and 
identified as tiny diamonds. Later silicon carbide (SiC) \citep{Bernatowicz_etal87}, graphite \citep{Amari_etal90}, corundum (Al$_{2}$O$_{3}$) 
\citep[Hutcheon et al.\ 1994;][]{Nittler_etal94}, 
silicon nitride (Si$_{3}$N$_{4}$) 
\citep[Russell et al.\ 1995;][]{Nittler_etal95} and 
spinel (MgAl$_{2}$O$_{4}$) \citep{Nittler_etal94} were identified. 
Some of the SiC and graphite grains have been found to carry small 
inclusions of Ti-, Mo- and Zr-carbides \citep{Bernatowicz_etal91}. 
 
Central to the identification of presolar grains is the determination of the isotopic composition of the 
grain and/or some trace elements trapped in the grains. As a rule the isotopic composition of the grain or some of its inclusions deviates strongly from the normal Solar System composition. The isotopic 
signatures of the grains contain information of the nucleosynthesis processes of the parent stars. 
Information on individual stars can be obtained by studying single grains by e.g.\ SIMS (Secondary 
Ion Mass Spectrometry) for the light to intermediate-mass elements, RIMS (Resonance Ionization Mass 
Spectrometry) for the heavy elements, and laser heating and gas mass spectrometry for He and Ne. TEM (Transmission 
Electron Microscopy) and SEM (Scanning Electron Microscopy) is used to study the crystal structure of the  
individual grains. 
 
Diamonds account for more than 99\% of the identified presolar meteoritic 
material, with an 
abundance that can exceed 0.1\% (1000\,ppm) of the matrix \citep{HussLewis95}, 
corresponding to more than 3\% of the total amount of carbon in the meteorite. 
Silicates are the second most abundant of the identified presolar grains. There seem to be two distinct groups of 
oxides and silicates, one originating from RGB stars and one from AGB stars. 
Identification of presolar silicates is complicated as they are ``diluted'' by 
the sea of ``normal silicates'' formed in the solar nebular, and analysis have therefore first begun to really develop with the new  
NanoSIMS technique \citep[e.g.][]{Hoppe_etal04} 
which allows an imaging search for isotopically anomalous phases {\it in situ} in the matrix of the meteorites.  
 
The best characterized of the presolar grains are SiC (6\,ppm) and almost all of the presolar SiC grains  
originate from AGB stars. Graphite (less than 1\,ppm) was traditionally assumed to originate from supernovae, 
but recent measurements of $s$-process signatures indicate that most high-density graphite grains are more 
likely to originate from AGB stars \citep{Croat_etal05}. 
\begin{table} 
\begin{center} 
\caption[]{Results of isotopic analyses on presolar grains. Data 
taken from \citet{Ott93} and \citet{HoppeOtt07}.} 
{\small 
\begin{tabular}{|l|c|c|c|c|}     \hline 
         &               &    &  &      \\ 
{\bf Mineral}  & {\bf Analysis} & {\bf Isotopic Anomalies} & {\bf Abund.} & {\bf Grain size}   \\  
   &   {\bf type}       &                 & {\bf ($\sim ppm$)} & {\bf ($\mu$m)}   \\ \hline 
Diamond  & Bulk          & Noble gases, N, Sr, Ba & 1500 & $0.002-0.003$  \\ \hline 
Silicon  & Single grain  & C, N, Si, Mg--Al, Ti, Ca, He, Ne & 30 & $0.1-30$ \\ 
Carbide  & Bulk          & Noble gases, Sr, Ba, Nd, Sm, Dy, Er &  &  \\  \hline 
Graphite & Single grain  & C, N, Mg--Al, O, Si, Ca, Ti, He, Ne & 1 & $0.1-10$ \\ 
         & Bulk          &     Noble gases & & \\ \hline 
Oxides    & Single grain  & O, Al--Mg, Ti  & 50 & $0.1-5$ \\ \hline 
Silicates    & Single grain  & O, Al--Mg, Ti  & 140 & $0.1-1$ \\ \hline 
Silicon  & Single grain  & C,N,Si & 1& 0.002 \\  
Nitride  &   &  & & \\ \hline 
\end{tabular} 
} 
\end{center} 
\end{table} 
 
Correlated measurements of as many elements as 
possible for a given grain are of special importance because such data makes it possible to set 
much tighter constraints on stellar sources and models of nucleosynthesis. 
In some cases, correlated measurements make it possible to construct 
stellar histories of individual grains \citep{Huss_etal97}. 

Observational data from both stars and presolar grains have clearly demonstrate the existence of ``cool bottom burning'' \citep{Wasserburg_etal95} occurring in low-mass RGB and AGB stars. 

Presolar grains hold great promise for improving our understanding of dust grain 
nucleation and growth in stellar environments, since formation processes are 
recorded in the detailed structures and compositions of the grains that can be 
characterized in minute detail. 
For instance, different types (masses/metallicities) of AGB stars seem to produce 
SiC grains with similar size distributions, which very likely 
reflects some common underlying mechanism.

Formation of grains that range in size from 100 nm or less to several microns 
seems to require a range of gas densities in outflows. In particular, formation 
of large ($>$1 $\mu$m) AGB grains on reasonable time-scales requires higher 
densities \citep{Nuth_etal06} than models indicate.  The density variations recorded by the grains 
could be caused by shocks in the outflows that are observed astronomically, and 
might highlight the need for two- and three-dimensional models of grain growth 
in outflows \citep{Bernatowicz_etal05}, or that pulsations will bring the materials
to favorable growth sites (sweet spots) where grains can stay for long enough time
periods to grow larger than the average grain size. 

As presolar grains seem to come from many stellar sources and to be relatively 
unprocessed, they also provide constraints on circumstellar dust production 
rates in the Galaxy. The relative abundances of the different types of presolar 
grain seem to indicate that AGB stars are the main dust producers in the Galaxy,
while supernovae only contribute a few percent \citep{Alexander97}.

%%%%%%%%%%%%%%%%%%
\section{Measuring optical properties of AGB dust analogues in the lab} \label{lab}
%%%%%%%%%%%%%%%%%% 
 
A particle placed in a beam of light will scatter some of the light 
incident on it (i.e.\ the light will change direction) and absorb some 
of the light (i.e.\ the electro-magnetic energy is transformed into 
other forms of energy by the particle). We say that the light has suffered extinction. 
The amount of light scattered and/or absorbed by the particle depends 
on the exact nature of the particle and also 
on the nature of the incident light. Consequently the extinction is 
dependent on the chemical composition of the particle and its size, shape 
and orientation, together with the wavelength and polarisation state of 
the light. The treatment of absorption and scattering of light by dust particles is a 
complicated problem within electromagnetic theory. For a comprehensive description see 
e.g.\,\citet{Bohren83}. 

\citet{Rayleigh71} developed the scattering theory for light scattered by dust particles with diameters smaller than the wavelength of the incident light, and showed that the amount of scattering is inversely proportional to the fourth power of the wavelength ($\lambda ^{-4}$). This means that the shorter the wavelength of the incident light, the more the light is scattered. When the dimensions of the dust particles increase, the $\lambda ^{-4}$ law ceases to be valid. Dispersion is then less selective with respect to $\lambda$, and Mie scattering theory \citep{Mie08} should be used. 
The complete formalism is sometimes referred to as Lorenz-Mie theory due to 
the previous work carried out by the Danish 
scientist Ludvig Lorenz \citep{Kragh91}. A complementary solution 
based on expansions of scalar potentials was given by \citet{Debye09}.
Mie theory is more generally valid and contains Rayleigh scattering as an 
approximation for particles small compared to the wavelength, and geometrical 
optics as an approximation for particles large compared with the wavelength. 
For sufficiently large dust particles, the dispersion of radiation approaches a $1/\lambda$ dependence, leading to diffuse reflection.  

There are no analytic solutions of the light scattering problem for 
particles of arbitrary shape, but in many cases, spectra of irregular 
particles can be approximated by a suitably averaging over different ellipsoidal 
shape parameters. With these approximations it is possible to obtain 
simple expressions for an average extinction cross section. 
 
The simplest approximation is that for spheres.  
Another approximation is a collection of randomly oriented 
ellipsoidal particles of all elliptisities (from spheres to infinite 
long needles). This continuous distribution of ellipsoids (CDE) was introduced by 
\citet{Bohren83}. 
\citet{Ossenkopf_etal92} have introduced a distribution with 
a higher probability for spheres (m-CDE), and being zero for the extreme values 
corresponding to infinitely thin needles or flattened pancakes. 
Calculations of the three different grain shape distributions, spheres, CDE 
and m-CDE for SiC grains can be seen in \citet{Mutschke_etal99}.
 
The two sets of quantities that are used to describe optical 
properties of dust particles are the real and imaginary parts of the complex refractive index 
$m = n + ik$ or the real and imaginary parts of the complex dielectric 
function (or relative permittivity) { $\epsilon = \epsilon^{\prime} + i 
\epsilon^{\prime \prime}$. 
These two sets of quantities are not independent; $\epsilon \equiv m^{2}$.  
The pair of quantities $n$ and $k$ are referred to as the optical constants, 
a bit misleading since they are frequency-dependent.  

For dust particles which are small compared to the wavelength, 
there are two distinct energy ranges in which 
resonances occur.  One is in the infrared, in the region of strong 
lattice bands between the transverse optical phonon frequency 
($\omega_{TO}$) and the longitudinal optical phonon frequency 
($\omega_{LO}$).  The other is in the ultraviolet and is due to 
the transitions of bound electrons. 
The optical properties corresponding to electronic transitions in condensed 
matter are in many cases 
described qualitatively correct by an oscillator model. 
The oscillator model is often used to derive $n$ and $k$ from reflection 
measurements of bulk materials. 
 
To obtain the optical properties of dust particles, there 
are two approaches. Either the optical constants can be determined by 
reflection and/or transmission measurements of bulk samples (e.g.\ single 
crystal, thin film or equivalent) and small-particle spectra can be 
calculated from these.  Or 
transmission measurements can be performed on samples of individual particles. Both 
methods have pros and cons. 
 
By determining $n$ and $k$ from bulk samples, it is assumed that the optical 
properties of the material can be completely specified. Therefore it is necessary to make 
assumptions of the grain size and morphology in-order to derive the optical 
properties of grains from the measurement of the particular material. 
Different assumptions about grain size and shape lead 
to very different spectral appearance.  
 
By measuring on samples of individual particles on the other hand, 
the grains have a certain size and morphology which of course 
should resemble what 
is expected of cosmic grains.  In space, grains will appear as single isolated 
particles in vacuum. To simulate this in the laboratory the grains are 
dispersed in a solid matrix which is transparent in the desired 
wavelength region. However, 
in the matrix there is a tendency of the grain sample to clump, 
which will result in a spectrum of small clusters of grains, which might 
be very different from a spectrum of single isolated particles. On top of 
this, the fact that the matrix has a refractive index different from 
vacuum will influence the band shape, see \citet{Mutschke_etal99} and 
\citet{Papoular_etal98} for a discussion of matrix effects. 
 
From a measurement of a particle sample, the extinction efficiency factor, 
Q$_{\rm ext}$, can be directly determined by a transmission measurement. 
In principle it is possible to determine $n$ and $k$ from the feature of a  
transmission (or absorption) measurement. But this is only a 
quantitative method {\it if} the path length is known and the medium is 
homogeneous. With the present method of obtaining 
laboratory spectra of particles embedded in a matrix, the assumption of 
a homogeneous medium will sometimes be far from reality. 
Clustering can cause a dramatic difference in the optical properties 
\citep{Huffman88}. 
 
This leaves the astronomer in a dilemma: Is it better to use $n$ and $k$ 
from bulk measurements or samples of individual particles included in a matrix measurement, 
when wanting to compare with observed astronomical spectra? There is no 
simple answer to this question and ``the solution'' will in many cases depend on the available 
laboratory data. The choice is to either use data of realistic grains, but 
possible agglomerated, or to use $n$ and $k$ from bulk 
measurements, for calculations of idealised grains in vacuum. 
A list of laboratory data (both 
bulk and particulate) can be found in the database by \citet{Henning_etal99}, 
where a broad collection of laboratory measurements from 
various scientific journals and handbooks has been compiled. 

\section{Dust formed in C-rich AGB stars} \label{carbon}

While carbon is expected to constitute a major fraction of the circumstellar
dust in carbon stars, its mineralogical form is still unclear.
Carbon has the unique property that the atoms can form three different
types of bonds through sp$^{1}$, sp$^{2}$ (graphite) and sp$^{3}$ (diamond)
hybridization. Amorphous carbon is a broad term covering materials which have 
a combination of the different bond types.

A number of observations of carbon-rich late-type stars indicate that amorphous carbon
is the dominant dust type in these stars \citep[e.g.\ Campbell et al.\ 1975; 
Sopka et al.\ 1985; Martin \& Rogers 1987;][]{Gurtler_etal96}. \nocite{Campbell_etal76} \nocite{Sopka_etal85} \nocite{MartinRogers87} 

Amorphous materials can show a whole range of different optical properties, related to
the exact micro-physical properties of the measured sample. Amorphous carbon is
an illustrative example of this, as the measured extinction can differ by a factor of 
10 (see Fig.\,\ref{aC}) depending on the detailed micro-physical properties of the
amorphous dust.

\begin{figure}[t]
\centering 
\leavevmode 
\epsfxsize=0.85 
\columnwidth 
\epsfbox{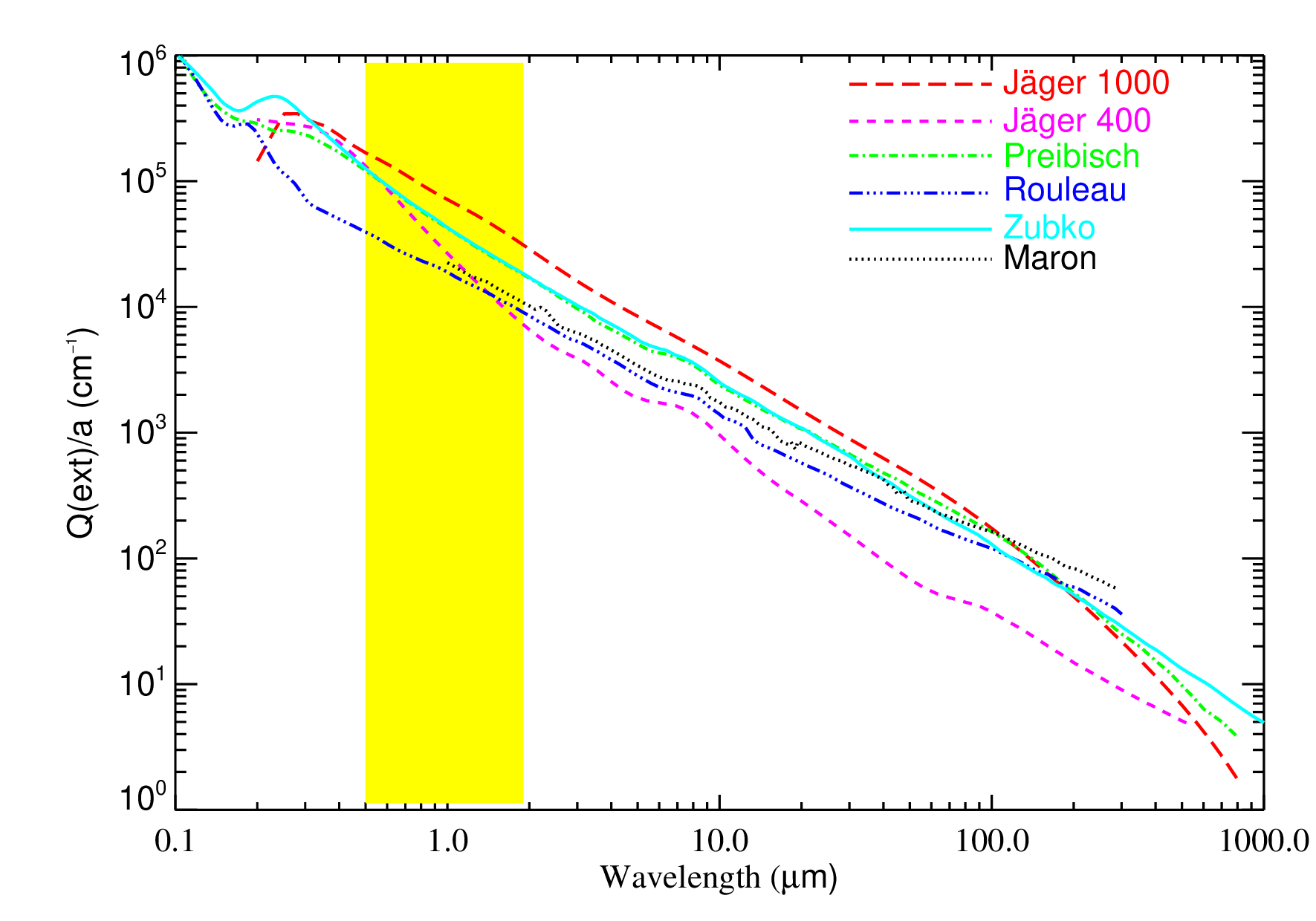} 
 \caption{Calculated extinction efficiency, $Q_{ext}$ over particle radius, $a$, of amorphous carbon from optical constants published  
 by \citet{Maron90}, \citet{RouleauMartin91}, \citet{Preibisch_etal93}, \citet{Zubko_etal96} and  
 two different samples (cel1000 and cel400) from \citet{Jager_etal98}. The J{\"a}ger~1000 sample contains significantly more $sp^2$ bonds than the J{\"a}ger~400 sample. The extinction efficiency factors 
 were calculated from the optical constants ($n$ and $k$) in the Rayleigh approximation for spheres. 
 The shaded area indicate the wavelength region of  maximum flux for typical AGB photospheres.  
 } 
\label{aC} 
\end{figure} 
 
Amorphous carbon grains are apparently a good candidate as the common
type of carbon grains present in circumstellar envelopes, but again the 
astronomer is left with a dilemma: Which of the available laboratory data should
one choose when they differ so much? Unless there is some indication from 
observations on which of the amorphous carbon laboratory data best represent the
stellar conditions, it isn't easy to say.  \citet{Andersen_etal03} have demonstrated
that although using one or the other of the amorphous dust data presented in Fig.\,\ref{aC}
didn't make a big difference for the hydrodynamic model structure, it had 
significant influence in the radiative transfer calculations and thereby influenced
the spectral energy distribution estimated for the circumstellar dust shell.

Thermodynamic equilibrium calculations performed by  
\citet{Friedemann69a,Friedemann69b} and \citet{Gilman69}
suggested that SiC particles
can form in the mass outflow of C-rich AGB stars.
The observations performed
by \citet{Hackwell72} and \citet{TreffersCohen74} presented the first observational  
evidence for the presence of SiC particles in stellar atmospheres.
A broad infrared emission feature seen in the spectra of many carbon stars, 
peaking between 11.0
and 11.5 $\mu$m is attributed to solid SiC particles and SiC
is believed to be a significant constituent of the dust around carbon
stars. 

Some years ago there was a lot of confusion as to whether the crystal structure of
SiC could be identified from spectra of C-stars, but laboratory measurements have 
now shown this is not possible, as the
crystal structure is not as significant as the morphology of the SiC grains. See \citet{Mutschke_etal99} for details.

One of the  main reasons for the huge mass loss of carbon-rich AGB stars seems to be the presence of 
newly formed dust grains. The strong shock waves in the stellar atmosphere cause a 
levitation of the outer layers. The cool and relatively dense environment which results 
from the levitation provides favorable conditions for the formation of molecules and 
grains.  Due to it high opacity and the resulting radiative pressure, 
dust plays an  important role in driving the wind.

\section{Dust formed in O-rich AGB stars} \label{oxygen}

\begin{figure}[t]
\centering 
\leavevmode 
\epsfxsize=0.85
\columnwidth 
\epsfbox{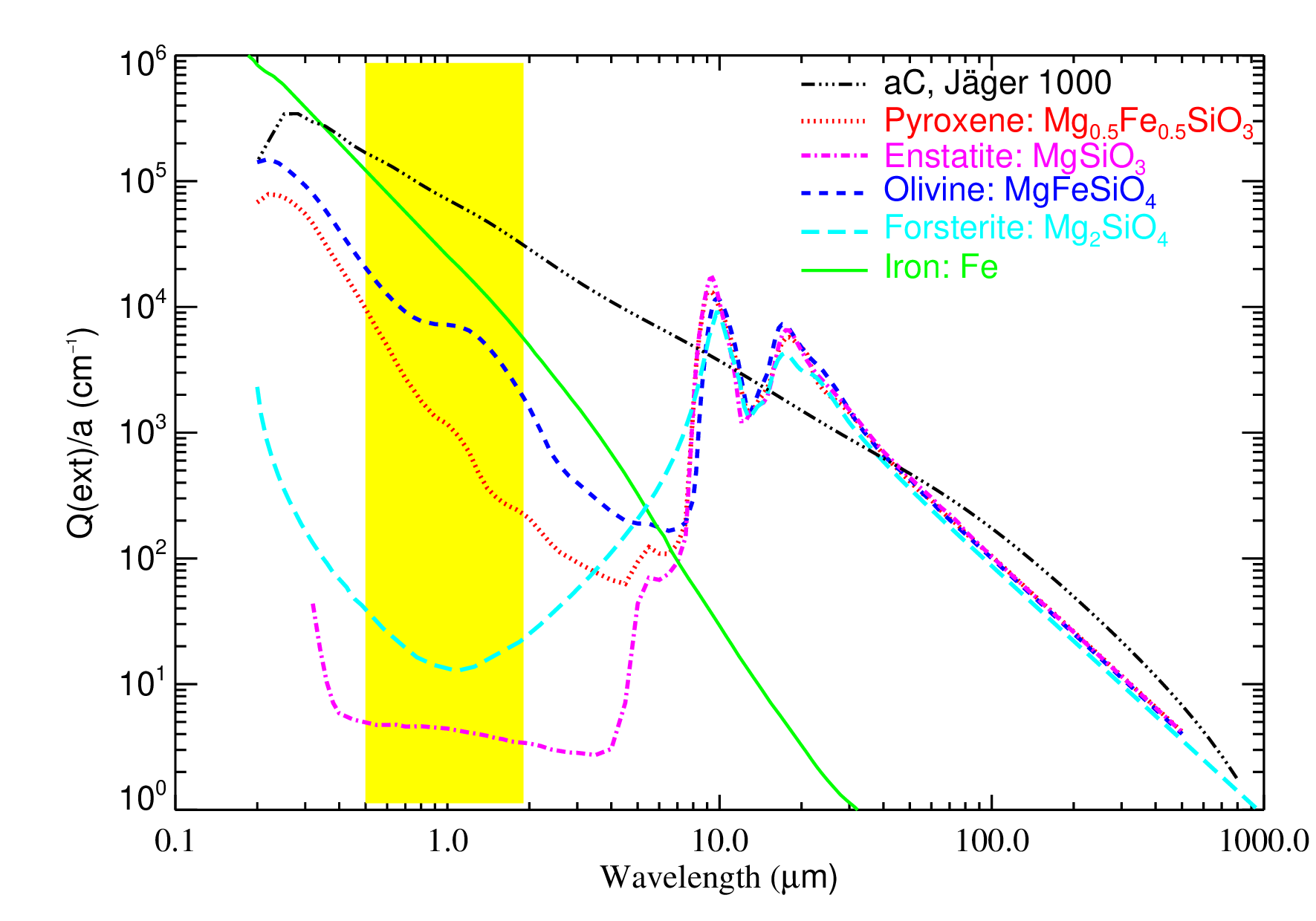} 
 \caption{Calculated extinction efficiency, $Q_{ext}$, over particle radius, 
 $a$, in the Rayleigh-limit for different types of silicates 
 \citep{Dorschner_etal95,Jager_etal03}. For comparison pure iron and amorphous carbon (cel~1000) from \citet{Jager_etal98} are shown. Notice that the different silicates have extinction efficiencies which are several orders of magnitude below that of amorphous carbon in the wavelength region of maximum flux for typical AGB photospheres (indicated by the shaded area). Iron rich silicates have stronger absorption around $1\mu$m that iron poor silicates. } 
\label{oxide} 
\end{figure} 
 
Magnesium-iron silicates have been found around evolved stars with an oxygen-rich dusty outflow \citep[e.g.\,Waters et al.\ 1996; Sylvester et al.\ 1999;][]{Molster_etal02}, \nocite{Waters_etal96} \nocite{Sylvester_etal99} as well as around stars with carbon-rich chemistry \citep[e.g.][]{Waters_etal98}. Only a fraction ($\sim 10$\%) of the observed silicates are in crystalline form \citep{Kemper_etal01}.

Silicates are the most stable condensates formed from the abundant elements 
O, Si, Mg and Fe. Out of these four elements silicate grains form as 
silica tetrahedras (SiO$_4$) combined with Mg$^{2+}$ or Fe$^{2+}$ cations. In the crystalline lattice structures it is possible for the tetrahedras to share their oxygen atoms with other tetrahedras and thereby form many different types of silicates \citep{MolsterKemper05},  which are described by 
\begin{center}
{Mg$_{2x}$Fe$_{2(1-x)}$SiO$_{4}$ with $x \in [0,1]$} \\ 
{ $x=1 \rightarrow$ forsterite, $x=0 \rightarrow$ fayalite, $0<x<1 \rightarrow$ olivine.} \\[0.3cm] 
{Mg$_{x}$Fe$_{(1-x)}$SiO$_{3}$ with $x \in [0,1]$} \\ 
{$x=1 \rightarrow$ enstatite, $x=0 \rightarrow$ ferrosilite, $0<x<1 \rightarrow$ pyroxene.} 
\end{center}

The optical properties of these silicates all have resonances around 10--20$\mu$m, due to the Si-O stretching and the O-Si-O bending mode arising from the silica-tetrahedras. Alignment of the tetrahedras may cause sharp peaked resonances, whereas amorphous silicates will show a broad feature which can be seen as a blend of such sharp resonances. 
The extinction efficiency of different silicates can be see in Fig.\,\ref{oxide}. For comparison the amorphous carbon data from \citet{Jager_etal98} (presented in Fig.\,\ref{aC}) and iron from \citet{Palik85} are also shown. What characterises silicate grains is the lack of strong absorption in the range where AGB stars have most of their flux. 

The slope of the dust absorption efficiency around the maximum flux of the 
AGB photosphere influences the dust grain temperature. The steeper 
the slope the higher the dust grain temperature will be. A high dust grain 
temperature will prohibit dust formation in the stellar atmosphere (see the contributions from H{\"o}fner and Woitke in these proceedings for more details). 
The iron rich silicate grains are therefore prevented from forming and the iron poor silicate grains do not have high enough opacity to drive the wind. These grain properties of silicate grains have currently turned the mass loss mechanism of oxygen-rich AGB stars into a bit of a puzzle.

\section{Summary}

Galaxies are constantly enriched by dust produced in the cool stellar atmospheres of AGB stars. Small amounts of these grains have been incorporated into presolar grains which can be extracted from meteorites. Results from isotopic studies of these grains have made it possible to trace the grains' to specific stellar sources. A major fraction of these grains originate from AGB stars at different phases in their AGB evolution. Given the precision of the laboratory isotopic analyses, which by far exceeds whatever can be hoped for in remote analyses, it is possible to obtain detailed information regarding nucleosynthesis and mixing in the parent stars as well as important hints towards obtaining a better understanding of the galactic chemical evolution. 

For interpretation of observations the single most important parameter for dust are the extinction properties of  a specific species. These are often measured in the laboratory on synthetically produced cosmic dust analogues.

\acknowledgements %%% Text of acknowledgements runs on after this command. 
The author would like to thank Susanne H{\"o}fner, Uffe Gr{\aa}e J{\o}rgensen and Peter Woitke for fruitful discussions and Franz Kerschbaum for endless patience.  
The Dark Cosmology Centre is funded by the Danish National Research Foundation.

%%% THE BIBLIOGRAPHY 
%%% 
%%% CONSULT SECTION 3 OF "INSTRUCTIONS FOR AUTHORS" FOR HOW TO USE NATBIB. 
%%% AUTHORS ARE ENCOURAGED TO USE EITHER THE "THEBIBLIOGRAPY" ENVIRONMENT 
%%% BY UNCOMMENTING (DELETING THE "%" SYMBOL) THE COMMANDS BELOW, OR BY 
%%% USING THE BIBTEX ENVIRONMENT. TO FIND OUT WHICH IS APPLICABLE TO YOUR 
%%% CONTRIBUTION, CONSULT THE VOLUME EDITORS FOR YOUR PROCEEDINGS. 
%%% 

\end{document}